\documentclass[sn-basic,Numbered]{sn-jnl}

\usepackage{graphicx}%
\usepackage{multirow}%
\usepackage{amsmath,amssymb,amsfonts}%
\usepackage{amsthm}%
\usepackage{mathrsfs}%
\usepackage{appendix}%
\usepackage{xcolor}%
\usepackage{textcomp}%
\usepackage{manyfoot}%
\usepackage{booktabs}%
\usepackage{algorithm}%
\usepackage{algorithmicx}%
\usepackage{algpseudocode}%
\usepackage{listings}%
\usepackage{tikz}

\theoremstyle{thmstyleone}%
\newtheorem{theorem}{Theorem}
\newtheorem{proposition}[theorem]{Proposition}%
\theoremstyle{thmstyletwo}%
\theoremstyle{thmstylethree}%
\usepackage{graphicx}
\usepackage{tikz}
\usetikzlibrary{positioning}
\usepackage{subcaption}
\usepackage{graphicx}
\usepackage{subcaption}
\begin{document}

\title[Article Title]{Revitalising the Single Batch Environment: A `Quest' to Achieve Fairness and Efficiency}


\author*[1]{\fnm{Supriya} \sur{Manna}}\email{supriya\_manna@srmap.edu.in}

\author[1]{\fnm{Krishna Siva Prasad} \sur{Mudigonda}}\email{krishnasivaprasad.m@srmap.edu.in}


\affil*[1]{\orgdiv{Dept. of Computer Science and Engineering}, \orgname{SRM University, AP}, \orgaddress{\street{Neerukonda-Kuragallu}, \city{Mangalagiri}, \postcode{522502}, \state{Andhra Pradesh}, \country{India}}}

\abstract{In the realm of computer systems, efficient utilisation of the CPU (Central Processing Unit) has always been a paramount concern. Researchers and engineers have long sought ways to optimise process execution on the CPU, leading to the emergence of CPU scheduling as a field of study. In this research, we have analysed the single offline batch processing and investigated other sophisticated paradigms such as time-sharing operating systems and wildly used algorithms, and their shortcomings. Our work is directed towards two fundamental aspects of scheduling: efficiency and fairness. We propose a novel algorithm for batch processing that operates on a preemptive model, dynamically assigning priorities based on a robust ratio, employing a dynamic time slice, and utilising periodic sorting to achieve fairness. By engineering this responsive and fair model, the proposed algorithm strikes a delicate balance between efficiency and fairness, providing an optimised solution for batch scheduling while ensuring system responsiveness.}

\keywords{CPU scheduling, Batch processing, Efficiency, Fairness, Greedy algorithm, Optimisation}



\maketitle

\section{Introduction}\label{sec1}

In the realm of computer systems, maximising the efficient utilisation of the CPU (Central Processing Unit) stands as a fundamental objective. From the earliest days of computing, relentless efforts by researchers and engineers have been devoted to devising sophisticated techniques aimed at optimising process execution on the CPU. The discipline of CPU scheduling has emerged as a response to this perpetual quest for enhanced efficiency and resource allocation. 

The roots of CPU scheduling \cite{silberschatz2006operating}, \cite{harki2020cpu} can be traced back to the early days of computing when batch processing systems were prevalent. In these systems, a set of fixed-size jobs was submitted as a batch to the computer, and the CPU had to execute them one after another. This approach suffered from inefficiencies as the CPU lacked responsiveness or processes waiting for the completion of others. Subsequently, to overcome certain limitations, scheduling algorithms were devised to manage the execution order of jobs, aiming to minimise idle time and maximise CPU utilisation. 

Efficient scheduling not only impacts system performance but also has significant economic implications. In today's digital era, where computational power is a valuable resource, optimising CPU scheduling can lead to substantial cost savings. By ensuring high throughput of the CPUs, organisations can complete computational tasks faster, reduce energy consumption, and ultimately save resources.

In addition to efficiency, the consideration of fairness in job selection is a crucial aspect of schedulers that is frequently under-addressed   \cite{ajtai1998fairness}. However, fairness is often viewed as a subjective metric lacking a universally agreed-upon definition in various task-scheduling contexts   \cite{wierman2011fairness}.

In this study, we return to the fundamental scheduling paradigm of batch processing. Here, we have a single, fixed-size queue of jobs, each with predetermined burst times. A single machine processes the queue, with preemption allowed between jobs. Despite the growing demand for advanced scheduling algorithms in recent times, this very foundational paradigm has comparatively been overlooked. As a result, the potential usage of the paradigm has also not been utilised across various domains   \cite{reuther2018scalable}. We acknowledge that to enhance the applicability of this paradigm in today's computing environment, we need to design algorithms that are not only succinctly efficient but also sufficiently fair   \cite{prabhakaran2014batch, robert2007non, zhang2020dybatch}. As most of the commonly used algorithms are primarily for time-sharing and multiprocessing systems   \cite{silberschatz2006operating}, the question of what feasible means to measure fairness and efficiency for this setting has not been effectively addressed. Our work addresses this question by first analysing these measures and then analyzing the classical algorithms commonly used in time-sharing and multi-programming systems   \cite{silberschatz2006operating} with respect to these measures to understand how they perform in terms of both efficient and fair distribution of job selections across a diverse set of job clusters. Lastly, through our efforts, we propose a novel algorithm that achieves both efficiency and fairness and outperforms traditional algorithms in striking a balance between these two factors. Our algorithm strives to enhance system productivity, reduce costs, and improve the overall user experience in various batch or batch-like computing environments. Specifically, our contributions are summarized as:
\begin{itemize}
\item We introduced a novel algorithm concerning fairness and efficiency in uni-processing batch environment;
\item We theoretically analysed and experimentally demonstrated its efficiency under a diverse and robust experimental setup and compared with other wildly used policies in the light of balancing efficiency and fairness;
\item We furthermore come up with an optimised version of our algorithm to reduce additional computational overhead.
\end{itemize}
Across the manuscript, we've interchangeably used the terms `Job', `Task', and `Process'. 

\section{Related Work} 
The primary goal of CPU scheduling is to allocate the CPU among multiple processes fairly and optimally. This involves making crucial decisions about process order and execution, taking into account various criteria and objectives. Key criteria for CPU scheduling include turnaround time, waiting time, response time etc. Balancing these criteria is essential to ensure fairness, efficiency, and responsiveness in the overall system performance. To achieve these goals, various CPU scheduling algorithms have been developed, each with its own advantages and trade-offs. The following section provides a comprehensive glossary of important terms and a detailed analysis of popular CPU scheduling algorithms, shedding light on their functioning and impact on system behaviour. By understanding the broader context and criteria of CPU scheduling, we can delve into the intricacies of different algorithms and evaluate their effectiveness in meeting the diverse needs of modern computing environments.

\subsection{Glossary}
\begin{itemize}

\item Arrival Time: The time at which a process arrives and becomes ready for execution.
\item Burst Time (\textit{bursttime}): The amount of estimated CPU time required by a process to complete its execution.
\item Remaining Time (\textit{remainingtime}): The amount of time still needed by a process to complete its execution. For example, if a process has a burst time of 10 seconds and has already been executed for 5 seconds, the remaining time would be (10-5)= 5 seconds.
\item Waiting Time (\textit{waitingtime}): The total time a process spends waiting in the ready queue before getting the CPU. For instance, if a process arrives at time 0, and there are two processes already in the queue, it would wait until the preceding process(es) leaves the CPU for its execution.
\item Turnaround Time (\textit{turnaroundtime}): The total time taken by a process to complete its execution, including both waiting time and execution time. This is also known as completion time which we've used interchangeably in the manuscript. Moreover, by definition \textit{turnaroundtime} = \textit{bursttime} + \textit{waitingtime}  \cite{silberschatz2006operating}. So, minimising \textit{turnaroundtime} is \textit{strategically equivalent} to minimising \textit{waitingtime}.  
\item Response Time (\textit{responsetime}): Response time, in the context of computing, signifies the period it takes for the CPU to react to a request initiated by a process. It essentially measures the interval between the arrival of a process and its initial execution.
\item Preemption: The act of interrupting the execution of a process before it completes its execution. Preemption allows for the allocation of CPU time to other processes with higher priority or in preemptive scheduling algorithms. Number of times it takes place has been referred as \textit{preemptioncount} in the paper.
\item Context Switching: The process of saving and restoring the state of a process so that it can be resumed from the same point when it is scheduled again. Context switching occurs during preemption or when a new process is selected for execution.
\item Quantum/Time Slice (\textit{here, timeQuantum}): The fixed amount of time allocated to a set of processes while scheduling (prevalent in Round Robin).
\item Priority: A value assigned to a process to determine its relative importance or priority in scheduling. Higher-priority processes are given precedence over lower-priority processes for CPU allocation.
\end{itemize}

Over time, researchers have proposed enhancements and improvements to classical algorithms (see Appendix A \ref{A}). For instance, variations of Round Robin, such as variations of Weighted Round Robin (WRR)  \cite{ji2003fair},  \cite{thiele2013improved} and Multilevel Queue Scheduling, have been introduced to address the limitations of strict time slicing. Additionally, policies like Priority Scheduling and Multilevel Feedback Queue Scheduling have been developed to incorporate process priority and dynamically adjust scheduling parameters.
Recent advancements in CPU scheduling have focused on adaptive and intelligent algorithms, leveraging machine learning and optimisation techniques  \cite{negi2005applying},  \cite{hicham2017comparative},  \cite{tehsin2020selection}. These algorithms aim to dynamically adapt to workload patterns, predict burst times, and improve system performance. Examples include Reinforcement Learning-based scheduling  \cite{lee2020panda},  \cite{fan2021dras} algorithms, fuzzy-logic based heuristics  \cite{alam2008finding},  \cite{butt2016novel},  \cite{alam2013fuzzy}  and Genetic Algorithm-based approaches  \cite{fleming2012new}. While these advanced frameworks for complex operating systems take huge additional computations, other strategies, like backfilling algorithms  \cite{rajaei2006comparison}, fair-share schedulers  \cite{kay1988fair}, gang scheduling  \cite{moschakis2012evaluation}, deadline-based schedulers  \cite{srinivasan2002deadline} etc are complex to implement, cost computational overhead and do not comprehend the inter-dependencies of different attributes in an easy manner and not always suitable and/or applicable to the batch processing. The wildly used algorithms in scheduling has been discussed in Appendix A (\ref{A}).

A batch processing system can be visualised as a fixed-size array of jobs. In this research, we have considered the case where the jobs have their predetermined \textit{bursttime}s. This is the most fundamental paradigm in scheduling theory which along with its several variants serves as the cornerstone for some of the most important industrial processes as well as in various domains in data-engineering  \cite{zhang2020dybatch}, big data and high-performance computing \cite{reuther2018scalable} etc. 
Now, our `\textit{Quest}' is to \textit{analyse} this very simple batch paradigm with the intention to \textit{investigate}, primarily the \textit{behaviour} of the two metrics researchers are mostly interested in:\textbf{ Efficiency} and \textbf{Fairness}.\newline\newline
\textsc{I. Efficiency}\newline
 Batch processing is heavily used in industrial setups  \cite{feng2021capacity, toksari2022single} where we are mostly concerned with the completion of all the jobs. In our work, there is no explicit deadline, penalty or priority of jobs. This is why, our primary means to quantify the efficiency is to measure the average waiting and turnaround time for the whole batch. Researchers, in the past, had also considered the number of \textit{tardy jobs}  \cite{chen2009minimizing}, \textit{maximum lateness}  \cite{hoogeveen1996minimizing}, and sometimes \textit{makespans}  \cite{reza2005flowshop} in a set of derivatives of the same paradigms which is not suitable and/or applicable for our current objectives of this work. \newline\newline

\textsc{II. Fairness}\newline

Efficiency in queuing systems generally has standard metric(s) and definitions to measure the performance of policies. Fairness, unlike efficiency, doesn't have a universally accepted metric to deal with \cite{wierman2011fairness}. Experts had mostly related fairness in terms of selection and providing proportionate \textit{timeQuantum} to jobs \cite{ajtai1998fairness, wierman2011fairness} in the past. However, to the best of our knowledge, there is no universal quantifiable metric of fairness for single-batch processing in a uni-processing system and the aforesaid improvements are also not primarily intended for fairness in our setting.  

In our study, our primary focus is on comparing algorithms based on their efficiency, i.e. in terms of minimizing average turnaround and waiting times as stated above. However, efficiency is not the sole consideration, as practitioners also emphasize the system's average response time \cite{gardner2017scheduling}, which is the duration from a job's arrival to its first response. In our context, where all jobs are present in the batch and no new jobs are added during processing, depending on the selection of the jobs at each iteration \textit{responsetime} provides an overview of how \textit{fairly} jobs have been selected. This is why, we quantify fairness by calculating the average \textit{responsetime} for the entire batch. For our algorithm, at each iteration, we calculate \textit{ response time} for all eligible jobs and then compute the mean of these \textit{responsetime}s when no jobs are left, generating the average \textit{ response time} for the whole batch.

To summarise, we evaluate algorithms using three key parameters: average \textit{waitingtime}, average \textit{turnaroundtime}, and average \textit{responsetime}. The average \textit{waitingtime} and average \textit{turnaroundtime} are indicators of efficiency, while the average \textit{responsetime} reflects the fairness of the algorithms. A competitive algorithm not only excels in efficiency but also maintains a maintains a satisfactory level of fairness. Lower turnaround and waiting times indicate higher efficiency, while a lower average \textit{responsetime} indicates fairer job selection.

\section{Proposed Work}
While empirically going through examples of process clusters, a very thoughtful observation for us was to discover the fact that any cluster of processes is never discrete throughout execution. In other words, the current attributes of a particular cluster of processes will inevitably change after some interval and we would be dealing with different types of sub-problems (here cluster of processes) in each cycle. In a multilevel queue or multilevel feedback queue for example, if we have an interval of ‘t’ units and a set of algorithms ‘S’, we may halt after each `t' unit of time, analyse the system, and can choose the ‘most suitable’ algorithm from ‘S’. For a tiebreaker amongst comparable algorithms, we can even take their space and time complexity or other parameters into account. Multilevel queues and multilevel feedback queues (MLQs/MLFQs) are computationally intensive  \cite{silberschatz2006operating} and can be overkill when striking a balance between efficiency and fairness, this very observation was the main inspiration to come up with a metric that can address the crucial aspects and intricacies of the process attributes after a certain period. Our approach is primarily \textbf{experimental} and \textbf{substantially empirical} in nature. After experimenting with several ways of measuring the changes in various parameters and their interdependencies, we have finalised the ratio in the algorithm below based on a few aspects elaborated in \ref{Analyis} section. While this may not be as sophisticated as MLQ / MLFQ, the interdependencies of attributes, if addressed effectively, lead to a near-optimal solution. With the potential to address the limitations of traditional batch processing, our approach opens doors to more conclusive approaches in this domain.

The name of our algorithm is \textit{FairBatch}.

\begin{algorithm}
    \caption{FairBatch(Batch)}
    \begin{algorithmic}
        \For {Process in Batch}
            \State{Process.\textit{remainingtime} $\gets$ \textit{bursttime}}
            \State{Process.\textit{waitingtime} $\gets$ 1}
            \State{Process.\textit{preemptioncount} $\gets$ 1}
        \EndFor
        \While {there is Process in Batch}
            \For {Process in Batch}
                \State{Process.\textit{fairnessRatio} $\gets$ (Process.\textit{bursttime} - Process.\textit{remainingtime} + Process.\textit{waitingtime}) / Process.\textit{bursttime} * Process.\textit{preemptioncount}}
            \EndFor
            \State{Sort Batch in non-decreasing order of \textit{fairnessRatio}}
            \State{\textit{timeQuantum}$\gets$ ceil((mean of \textit{remainingtime} of processes + median of \textit{remainingtime} of processes)/2);}
            \While{\textit{timeQuantum} lasts}
                \For{Process in Batch}
                    \State{Run Process for $\delta$ unit of time}\Comment{$\delta$ $\gets$ $\min(\textit{timeQuantum}, Process\textit{.remainingtime})$}
                    \State{\textit{timeQuantum} $\gets$\textit{timeQuantum-$\delta$}}
                    \State{Process.\textit{preemptioncount}$\gets$Process.\textit{preemptioncount}+1}
                    \State{Process.\textit{waitingtime}$\gets$Process.\textit{waitingtime} + $\sum$\textit{remainingtime} for proceeding Process(es)}
                    \State{\mbox{Process.\textit{remainingtime}$\gets$Process.\textit{remainingtime}-$\delta$}}
                    \If{Process.\textit{remainingtime}=0}
                        \State{Remove Process}
                    \EndIf
                \EndFor
            \EndWhile
            \For{Process(es) didn't execute}
                \State{Process.\textit{waitingtime}$\gets$Process.\textit{waitingtime}+\textit{timeQuantum}}
            \EndFor
        \EndWhile
    \end{algorithmic}
\end{algorithm}

 \begin{figure}
    \centering
        \includegraphics[width=0.7\textwidth]{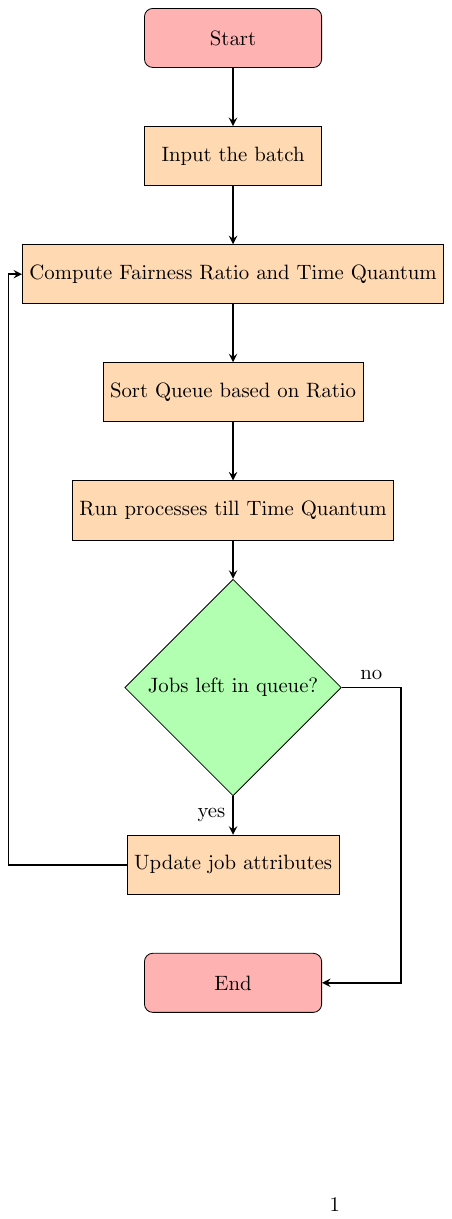}
        \caption{Flowchart of the algorithm}
    \end{figure}
    
\subsection{Analysis}\label{Analyis}
The proposed algorithm starts with calculating the ‘\textit{fairnessRatio}’ and sorting the processes based on the calculation. The aspects we have considered while designing the ratio are elaborated in following points. \newline\newline
\begin{equation} \label{one}
\textit{fairnessRatio}= \frac{bursttime - remainingtime + \textit{waitingtime}} {bursttime * preemptioncount}
\end{equation}

\textsc{I. Balanced Selection of Jobs}\newline

In equation \ref{one} the ratio tracks both how much a process has been waiting before execution, and how much progress has been made, before scheduling. In the numerator, \textit{(bursttime - remainingtime)} is the measure of the amount of \textit{bursttime} which has been consumed till a particular iteration, or in other words, this is the progress\ref{progress} of a process till the current iteration. The \textit{waitingtime} is a measure of the total amount of time one process has waited till the time it gets under execution. 

As the processes, if not under execution, have to wait for the whole time quanta (\textit{timeQuantum}) allotted for a particular cycle before moving to next iteration, the update at each cycle would be:

\begin{equation}
\text{p.waitingtime := p.waitingtime + timeQuantum}\quad\forall p \in W 
\end{equation}

W is the set of processes that didn't execute in a particular iteration, p is a process, \textit{p.waitingtime} is the \textit{waitingtime} of p.

The proposed algorithm takes a unique approach by considering \textbf{both shorter and longer processes} (or part of processes) to ensure a \textit{balanced} and \textit{efficient} execution. By using the \textit{fairnessRatio} and sorting mechanisms, the algorithm strategically selects a set of jobs that allows \textbf{shorter jobs to make more progress} in each iteration, while also \textbf{reducing the waiting time for longer processes}. As a result, the algorithm achieves a well-rounded mix of processes, promoting fair distribution of CPU resources and significantly improving the response time of longer processes. The approach's dynamic nature ensures that the scheduler efficiently handles various process types, creating an optimal balance between fairness and efficiency.

To address a specific edge case in the algorithm, we initialise the \textit{waitingtime} with one. This situation arises when a large process arrives at the front of the queue in the very first iteration, followed by significantly smaller processes. In the first cycle, the \textit{remainingtime} and \textit{bursttime} of processes is equal. If both the (\textit{bursttime - remainingtime}) and \textit{waitingtime} in eq.\ref{one} are being initialised with 0, it will act as simple fcfs scheduling which makes subsequent processes waiting for execution from the very first cycle, negatively impacting the overall efficiency of the whole batch. Alternatively, we can add 1 to \textit{waitingtime} if we initialised it to 0. Considering other attributes not changing, \textit{waitingtime} time will be increasing the LHS  of eq.\ref{one}  proportionately with respect to its magnitude.\newline

\textsc{II. Progress in Fairness ratio}\label{progress}\newline

\textit{Progress} is another pivotal dimension that affects fairness. Processes that have made substantial \textit{progress} in their execution should be prioritised to avoid unnecessary interruptions and context switches. By incorporating the \textit{progress} made by each process, as calculated by the difference between \textit{bursttime} and \textit{remainingtime} in the numerator of \textit{eq1}, the \textit{\textit{fairnessRatio}} acknowledges the importance of honouring the advancements of processes. The more the \textit{progress}, the lesser the \textit{remainingtime}. Under a fixed \textit{timeQunatum}, shorter processes are given more opportunities to make \textit{progress} due to their shorter \textit{bursttime}. This is achieved by inversely weighting the \textit{bursttime} and \text{remainingtime} in the ratio calculation. Considering other attributes not changing, \textit{progress} will be increasing the LHS  of eq.1  proportionately with respect to its magnitude.\newline

\textsc{III. Limiting Preemption using Fairness ratio}\newline

The \textit{fairnessRatio} takes into account the \textit{preemptioncount} (initialised with 1 to remove unnecessary Zero-Division-Error or explicitly 1 can be added if initialised as 0 ), which considers the cost of context switching and interrupts. By inversely weighting the \textit{fairnessRatio} based on the \textit{preemptioncount}, the algorithm promotes efficiency by reducing unnecessary preemption, minimising CPU overhead, and enhancing responsiveness.



\begin{proposition}
FairBatch works with at least 1 process
\end{proposition}
\begin{proof}
    \begin{enumerate}
\item Assume that there are no processes in the queue ($|P| = 0$, $P$: set of processes in the queue).
\item In this case, the while loop condition $|P| > 0$ is not satisfied.
\item According to Algorithm 1, if the while loop condition is not satisfied, the algorithm won't run.
\item Therefore, if $|P| = 0$, the algorithm won't run.
\end{enumerate}

This proves that the presence of at least one process in the queue is necessary for the algorithm to run. Hence, the proposition is proven.

\end{proof}
\begin{proposition}
    There is no preemption within processes in a particular iteration in FairBatch
\end{proposition}
\begin{proof}
   Consider the FairBatch algorithm (Algorithm 1) with a fixed time quantum denoted as \(T\). The algorithm operates as follows:

\begin{enumerate}
\item The algorithm runs until the time quantum \(T\) is exhausted.
\item During this time, it sequentially processes the tasks from the front of the sorted queue.
\item If \(T\) is less than the remaining time of the first task in the queue, only a proportional fraction of that task's execution is performed in the current iteration.
\item After the complete execution of a task or its proportional fraction, if there is time remaining in \(T\), the algorithm proceeds to the next task in the queue.
\item This process continues until \(T\) is exhausted without reordering the queue or allowing other processes to interrupt a currently executing process.
\end{enumerate}

To prove the absence of preemption within processes in a particular iteration, we will employ a proof by contradiction. 

Assume, for the sake of contradiction, that there is preemption within processes in a particular iteration in FairBatch. Let \(T\) be the fixed time quantum for this iteration. We define the following terms:

\begin{itemize}
\item \(P_i\) represents the \(i\)-th process in the queue.
\item \(R_i\) represents the remaining time of process \(P_i\) at the beginning of the iteration.
\item \(E_i\) represents the execution time allocated to process \(P_i\) within the iteration.
\item \(Q\) represents the queue of processes at the beginning of the iteration.
\end{itemize}

Now, consider the scenario in which preemption occurs within the iteration:

\begin{enumerate}
\item Process \(P_1\) begins execution with remaining time \(R_1\) and is interrupted after \(E_1\) time units (where \(0 < E_1 < R_1\)).
\item The queue \(Q\) is reevaluated, and other processes may be given a chance to execute.
\item After \(P_1\) is preempted, it is possible that another process \(P_j\) with \(j > 1\) starts executing.
\end{enumerate}
Therefore, this scenario of preemption within an iteration contradicts the fundamental principles of the FairBatch algorithm. Hence, we conclude that there is no preemption within processes in a particular iteration in \textit{FairBatch}.

\end{proof}
\begin{proposition}
    Between 2 consecutive iterations, there is at max 1 preemption.
\end{proposition}
\begin{proof}
   Let \(i\) represent the current iteration, and \(i+1\) the next iteration. We will analyze the occurrence of preemptions.

1. If the \(i\)-th iteration completes with all processes being completely executed, the next iteration begins with a different set of processes. In this scenario, there is no possibility of preemption.
   
2. Now, consider the case where not all processes are completely executed in the \(i\)-th iteration.

   a. Let \(P_k\) be the last process executed in the \(i\)-th iteration with the highest fairness ratio. This process may still have the highest fairness ratio in the \((i+1)\)-th iteration.

   b. There are two cases to consider:

      - \textbf{Case 1}: If \(P_k\) is chosen for execution at the beginning of the \((i+1)\)-th iteration.
        - In this case, there is no preemption in a continuous frame of reference assuming no delays between the iterations.

      - \textbf{Case 2}: If a different process \(P_j\) is chosen at the very front of the queue in the \((i+1)\)-th iteration.
        - This results in a preemption.

Therefore, in either case, there is either no preemption or at most one preemption between two consecutive iterations.

This completes the proof, demonstrating that between two consecutive iterations in FairBatch, there can be at most one preemption.\newline\newline
\end{proof}

\subsection{Further Analysis of the Parameters:}

In our investigation, we will analyze when our setting achieves the minimum and maximum waiting and response time.

\begin{itemize}

\item{Minimal Average Response Time}:\newline

Under our current setting, out of n jobs in a batch, only 1 can be under execution at a time and rest `n-1' job wait till there is any context switching. The greedy way is allocating each task the minimum possible execution time, regardless of its \textit{bursttime} and moving on to the next processes. For example, if there are `n' number of tasks and if each task is granted `k' unit of time, the first process experiences no initial delay, the second incurs a k-unit wait due to the first task, and so on, with the n-th process waiting for k unit for each preceding task. In this scenario, the average response time is calculated as:
\[
\frac{0 + k + 2k + \ldots + k(n-1)}{n} = \frac{k.(n-1)}{2}
\]
\textbf{remark:} Usually under general settings, k = 1 \cite{silberschatz2006operating}.\newline

\item{Minimal Average Waiting Time}:\newline

It is well-known that the SJF/SRTF (preemptive version of SJF) Rule minimises the average waiting time.  \cite{conway1967theory, smith1978new}

\begin{proposition}
    In our setting, SJF and SRTF are equivalent in terms of working and performance
\end{proposition}
\begin{proof}
    SRTF switches the context iff at any instance it discovers some lower burst(s) compared to the process currently under execution. In our setting, we have all the processes in the batch available from the very beginning and there is no addition of other jobs at any time. SRTF starts with the least burst from the batch and after each unit of time, the remaining time of the least burst is trivially the least (as there is guaranteed progress using proposition 3.1). This is why there is no preemption as there cannot be any other job at any instance which would have a lesser burst than the currently executing process for all jobs in the batch. This is exactly the way SJF works.  
\end{proof}

\textbf{remark}: For the sake of consistency, we have used SRTF and SJF interchangeably throughout the manuscript.\newline

\item{Maximum Average Response Time for the Batch}:\newline

 In our setting, the batch will experience the maximum response time when all the waiting processes have to wait the most. Exactly `n-1' jobs from a batch with `n' jobs will wait whenever exactly 1 job is under execution. If every time the currently executing job itself is of the highest burst amongst the processes and there is no preemption until the entire completion of the executing job, the rest of the jobs wait for the longest possible time. Longest Process First(LPT) chooses the longest job to execute without preemption. As a result, in our environment for a single batch with a single processing unit, it trivially maximises the \textit{responsetime}. It also maximises the average completion time \cite{conway1967theory}. 

\item{Maximum Average Waiting Time for the Batch}:\newline

 LPT is the inverse of the SJF policy and maximises mean completion time \cite{conway1967theory}. However, since the conventional approach prioritizes minimization of completion time, especially in our comparison with SRTF/SJF algorithms, we opt to evaluate the preemptive variant, LRTF. This choice aims to assess its potential for enhanced response times due to its aggressive preemptive nature.\newline

\end{itemize}
\textbf{remark:} As our main goal is to \textit{revitalise} the single batch environment, we've given the results for the \textit{extreme} cases above but in realistic scenarios, to the best of our knowledge, there is no single algorithm that simultaneously achieves the best of 2 world (lowest possible average waiting and response time). Any algorithm including \textit{FairBatch}, designed for this setting, would always fall within the extreme ranges demonstrated. As the concentration for this research is on considering both waiting, turnaround \textbf{and} response time \textbf{together} across the distribution, the trivial case for lowest both waiting and response time is achieved when simply the batch has 1 job using proposition 3.1. However algorithmic behaviour changes and shows different alignments with regards to the underlying distribution of burst time of the jobs.  

This is why,  we've studied and investigated exhaustively how our algorithms \textit{behave} under our robust experimental setup in Section 4. Now, there can be many types of distributions, for our research, we have considered those that are commonly observed in real-world scenarios the most and can simulate the problems that we had mentioned earlier such as convoy and starvation effects etc.

\subsection{Importance of a Suitable Time Quantum}

In a queue of processes, even when executing just one process for the minimal feasible time, (say ``t") for the sake of achieving utmost fairness, we must re-calculate the fairness ratio of every process within the batch after each ``t" unit of time.  Now ``t" mathematically can be arbitrarily small or large. If ``t" is exceedingly small, process progress becomes exceedingly marginal with excessively unnecessary context switches, while an arbitrarily large ``t" renders the algorithm's behaviour akin to First-Come-First-Serve (FCFS) for the majority of the time -- both of which contradicts the philosophy of FairBatch. Moreover in realistic scenarios, the distribution on job's \textit{bursttime}s may follow several distribution or might be completely random in nature. For example, in a positively skewed distribution, the mean is greater than the median and visa versa for negatively skewed counterpart. Regardless of distributions or random distributions, mean and median are always prominent factors to get an overview of the central tendency. This is why we have diligently explored diverse statistical formulations across current literatures  \cite{mora2017modified, shyam2014improved, hyytia2016round, xiuqin2012diffserv, banerjee2012comparative, mishra2014improved}, and have finalised,  inspired by the empirical work presented in  \cite{sharma2022new}.

\subsection{Optimisation of the Algorithm}
We can reduce the computational overhead produced by sorting the whole batch at each cycle by employing an efficient framework for grouping up and running the processes sequentially. 

- Consider the following for a particular cycle :

\begin{itemize}
\item \textit{S} is the set of available processes with remaining time \(> 0\), \(|S| = n\)
\item \textit{R} is the set of \textit{fairnessRatio}s \(f_i\) of processes.
\item \textit{timeQuantum} = \(W\).
\item \(r_i\) is the \textit{remaining time} of process \(i\) in \(S\).
\end{itemize}

\begin{algorithm}
    \caption{Select(\(R, W, S\))}
    \begin{algorithmic}
        \State Returns a set of chosen processes (\(s\))
        \State Compute the median \(m\) of \(R\)
        \State Determine:
        \State \(R1 = \{f_i \mid f_i > m, i \in S\}\), \(W1 = \sum_{i \in R1} r_i\)
        \State \(R2 = \{f_i \mid f_i = m, i \in S\}\), \(W2 = \sum_{i \in R2} r_i\)
        \State \(R3 = \{f_i \mid f_i < m, i \in S\}\), \(W3 = \sum_{i \in R3} r_i\)
        
        \If{\(W1 > W\)}
            \State Recursively take processes from \(R1\) in \(s\)
            \State Return the complete solution
        \Else
            \While{\(W \neq 0\) and \(R2 \neq \emptyset\)}
                \State Add processes from \(R2\) to \(s\)
            \EndWhile
            \If{\(W = 0\)}
                \State Return processes in \(R1 \cup R2\) in \(s\)
            \Else
                \State \(W \gets W - (W1 + W2)\)
                \State Recursively take processes from \(R3\) in \(s\)
                \State Return processes in \(R1 \cup R2\) and processes from the recursive call
            \EndIf
        \EndIf
    \end{algorithmic}
\end{algorithm}

\begin{algorithm}
    \caption{Runner($W, s, S$)}
    \begin{algorithmic}
    \State Sort s in non-decreasing order of fairnessRatio of processes
    
    \While{$W != 0$}
         \For{Process in Batch}
                    \State{Run Process for $\delta$ unit of time}\Comment{$\delta$ $\gets$ $\min(\textit{timeQuantum}, Process\textit{.remainingtime})$}
                    \State{\textit{timeQuantum} $\gets$\textit{timeQuantum-$\delta$}}
                    \State{Process.\textit{preemptioncount}$\gets$Process.\textit{preemptioncount}+1}
                    \State{Process.\textit{waitingtime}$\gets$Process.\textit{waitingtime}+ $\sum$\textit{remainingtime} for proceeding Process(es)}
                    \State{\mbox{Process.\textit{remainingtime}$\gets$Process.\textit{remainingtime}-$\delta$}}
                \If{Process is completed}
                    \State{S $\gets$ S - Process}
                \EndIf
            \EndFor
        \EndWhile
    \end{algorithmic}
\end{algorithm}

This `Select' algorithm outputs a set of appropriate proportions of job(s) based on \textit{fairnessRatio}s and \textit{timeQuantum}. The `Runner' algorithm takes the set of processes `s' as a input from `Select'. It sorts the processes in `s' based on their \textit{fairnessRatio}. Run the processes sequentially and update their attributes. After the cycle ends, the attributes of the process not were under execution is updated and the scheduler runs the subsequent cycles till there is process left in the batch.\newline
The `Select' algorithm starts by taking as many processes with values greater than the median as possible, staying within the weight constraint. If there is any leftover capacity, it takes processes with values equal to the median. If further capacity remains, the algorithm recursively considers combinations of processes to maximize the total value while respecting the \textit{timeQuantum} limit. The median calculation, utilized in both \textit{FairBatch} and the `Select' procedure, can be computed in linear time. In `Select', each recursive call requires linear time, excluding the time spent on potential recursive calls it may make. Since there is only one recursive call, it pertains to a problem size at most half of the original. As a result, the running time can be expressed by the following recurrence relation:
T(n) $<=$ T(n/2)+ $\theta$(n) therefore, T(n) = O(n), using master’s theorem.\newline
If the `Select' algorithm returns `k' jobs to be scheduled, `Runner' will take additional k.log k asymptotic time complete the scheduling due to sorting. 
So overall the scheduler takes: O(n+ k.log k) per cycle where k $<=$ n.

\section{Experimental Setup}

 FCFS is the most commonly used algorithm for batch jobs but we've also taken preemptive and advanced algorithms like SRTF, LRTF, RR, and CFS along with FCFS for comparison. Along with 
 theoretically well-grounded policies such as SRTF, LRTF, RR and FCFS, we have considered adopting the implementation CFS algorithm\footnote{\url{https://elixir.bootlin.com/linux/v5.19.9/source/kernel/sched/fair.c\#L7429}} for a uni-processing system as per our setting\footnote{following a similar pythonic adaptation used:\url{https://github.com/SanchithHegde/completely-fair-scheduler} }.  We consider the default `nice' value of individual jobs in the batch as 0\footnote{\url{https://man7.org/linux/man-pages/man7/sched.7.html}}. By comparing with these traditional and contemporary scheduling approaches enriches the evaluation, blending theoretical robustness with practical relevance, and ensuring a comprehensive assessment.

\begin{figure}
  \centering
  \begin{tikzpicture}
    \node[inner sep=0pt] (exp) {\includegraphics[width=\textwidth]{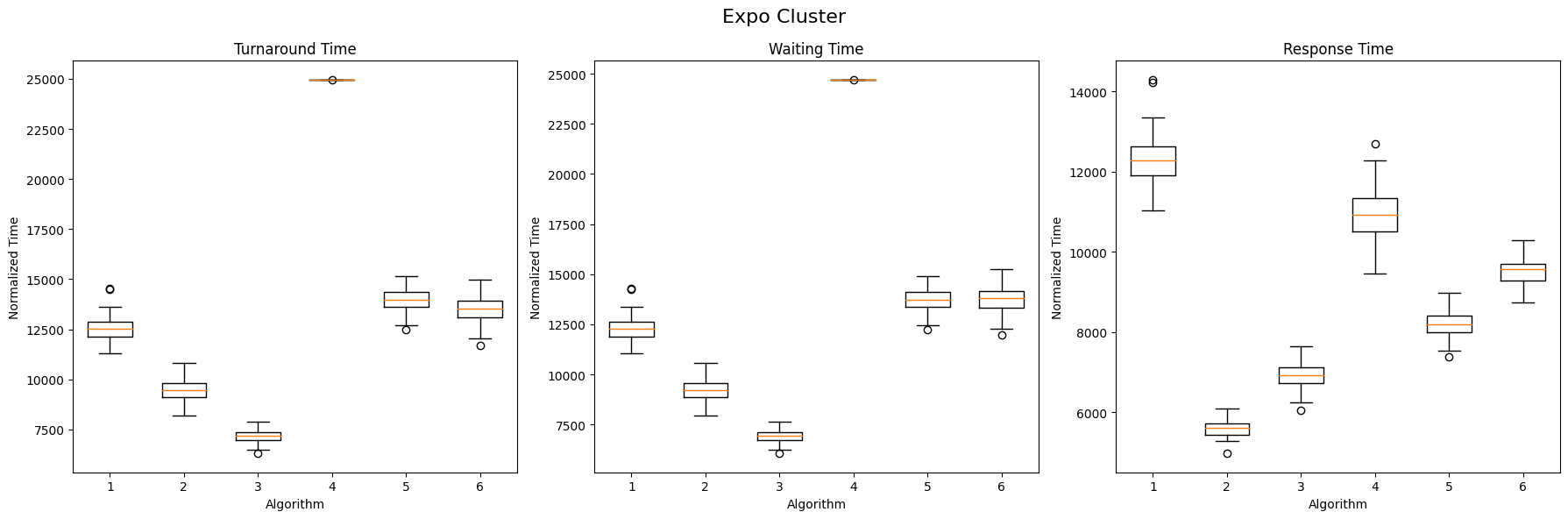}};
    \node[inner sep=0pt, below=0.2cm of exp] (geom) {\includegraphics[width=\textwidth]{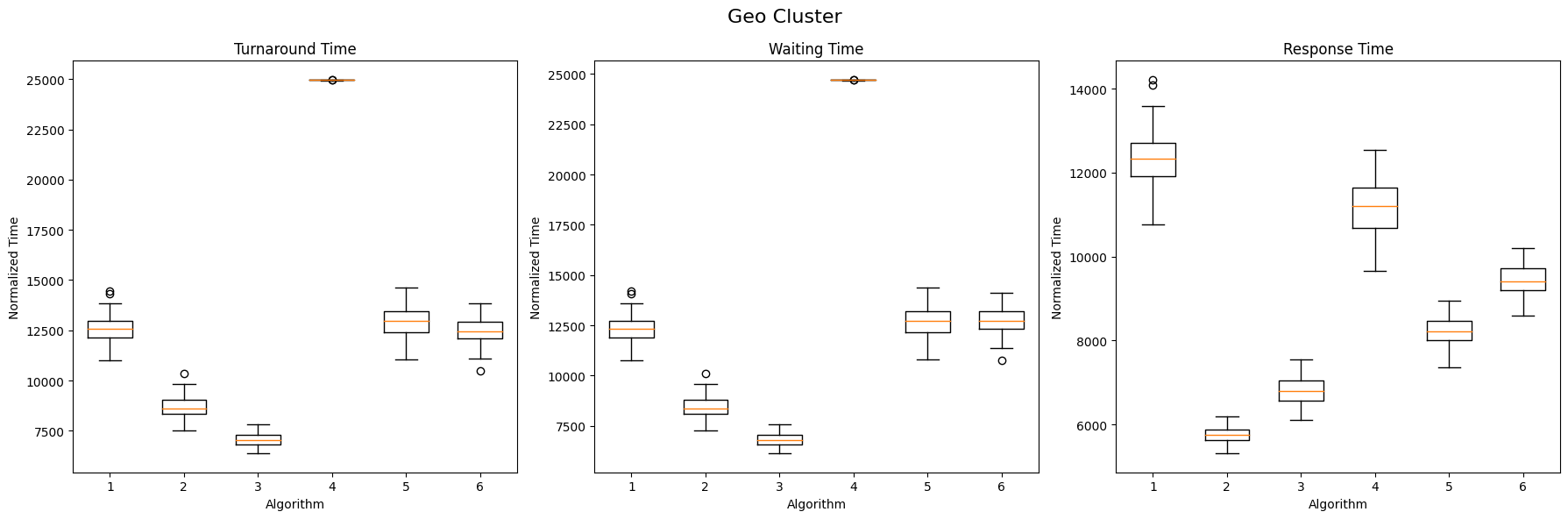}};
    \node[inner sep=0pt, below=0.2cm of geom] (negbin) {\includegraphics[width=\textwidth]{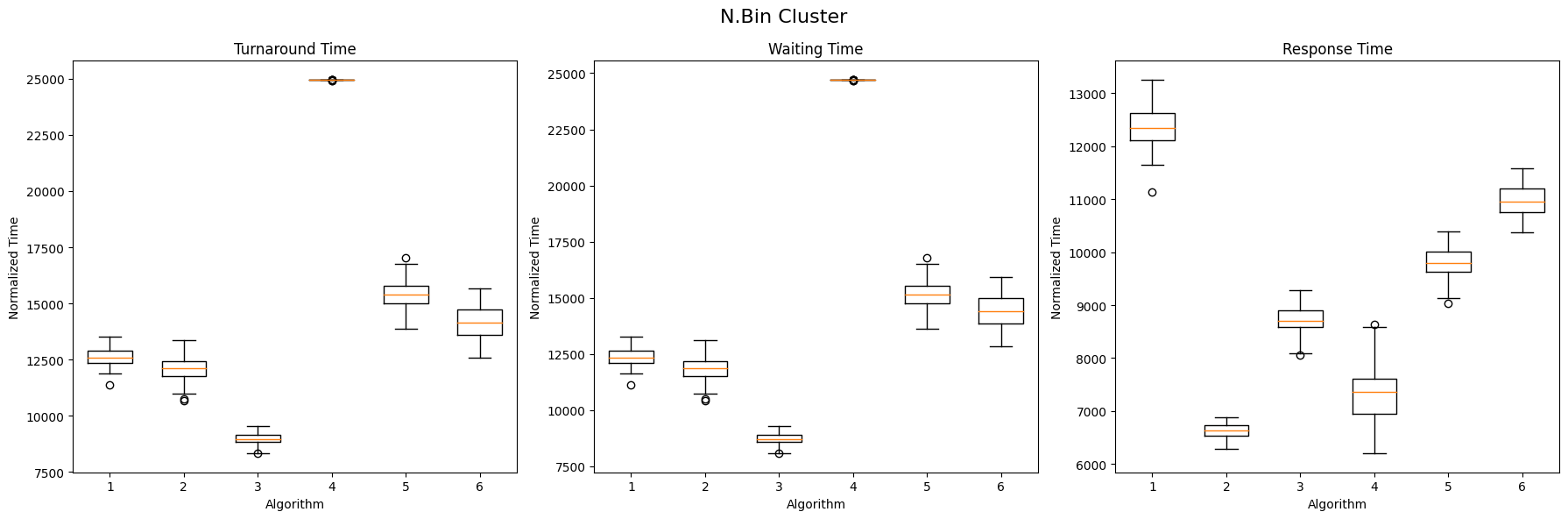}};
    \draw[black, line width=0.5pt] (exp.north west) rectangle (negbin.south east);
  \end{tikzpicture}
  \caption{Exponential, Geometric \& Negative Binomial clusters}
  \label{fig:2}
\end{figure}

\begin{figure}
  \centering
  \begin{tikzpicture}
    \node[inner sep=0pt] (exp) {\includegraphics[width=\textwidth]{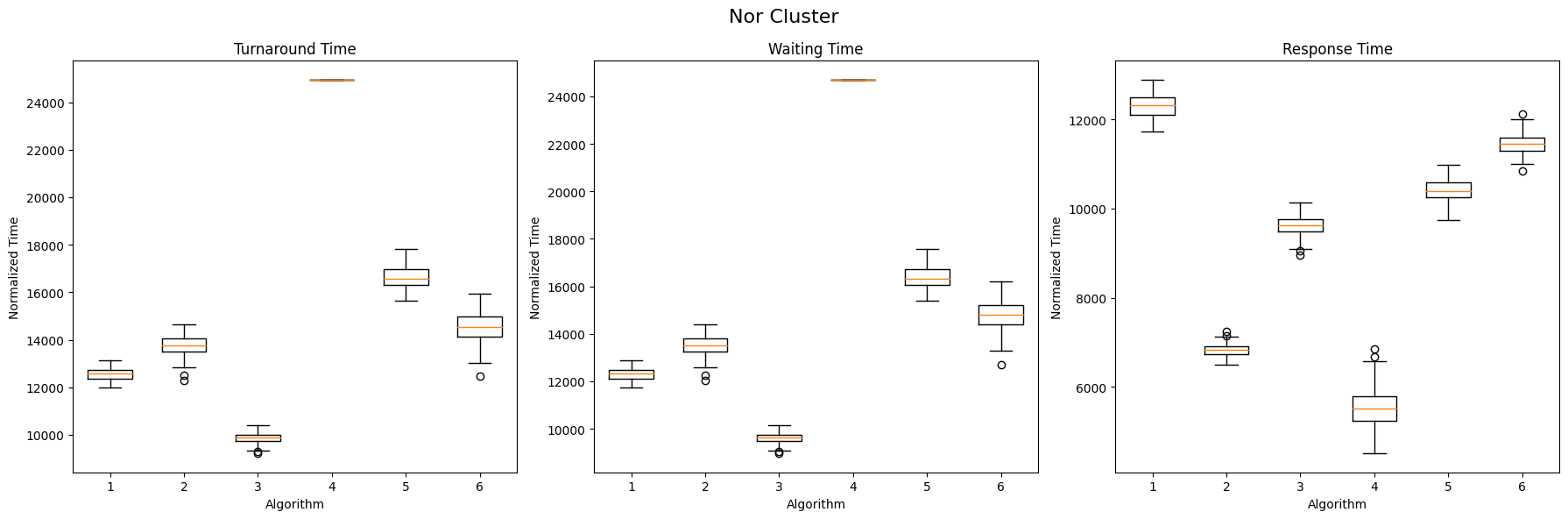}};
    \node[inner sep=0pt, below=0.2cm of exp] (geom) {\includegraphics[width=\textwidth]{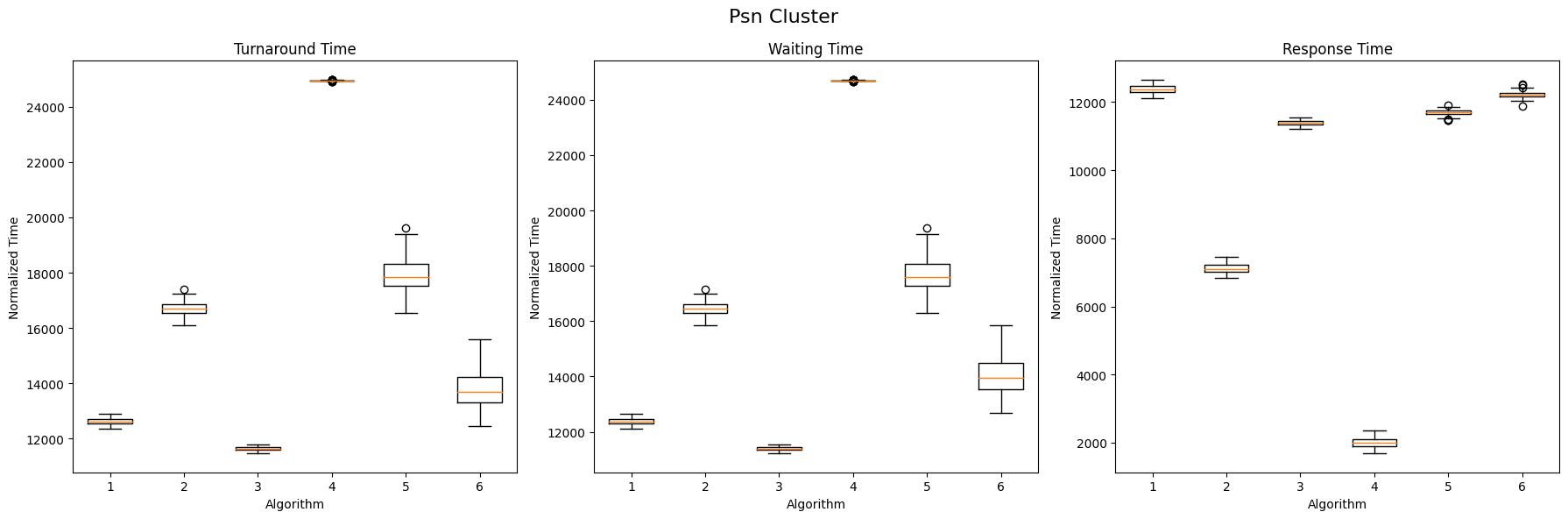}};
    \node[inner sep=0pt, below=0.2cm of geom] (negbin) {\includegraphics[width=\textwidth]{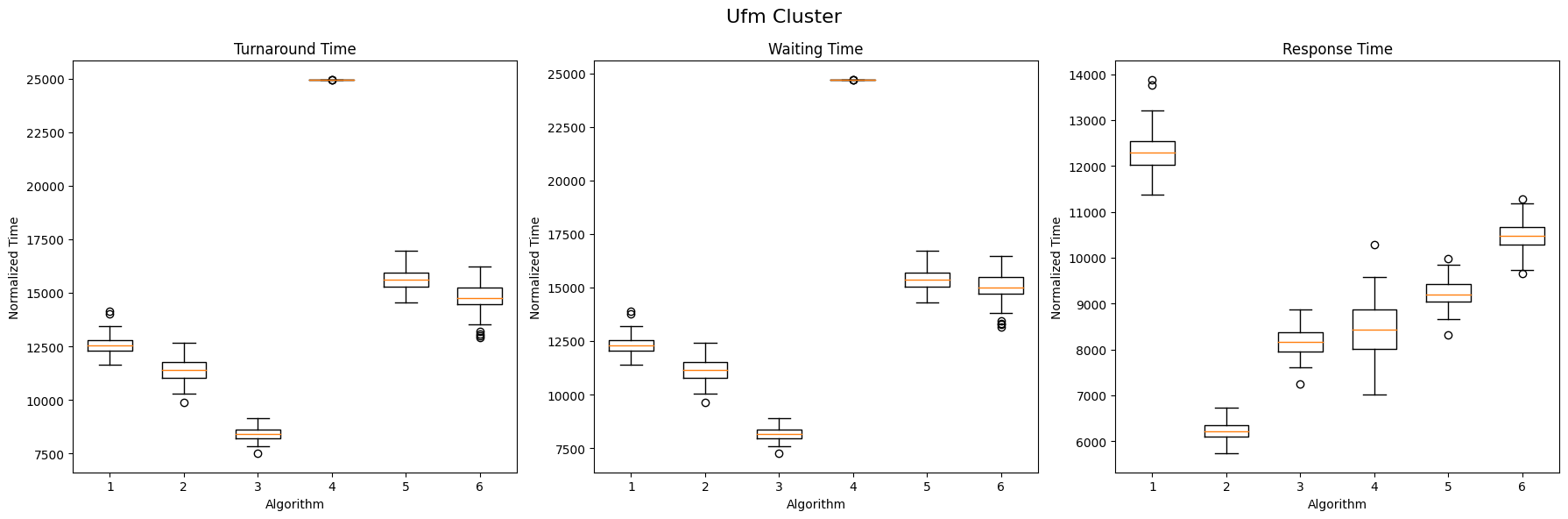}};
    \draw[black, line width=0.5pt] (exp.north west) rectangle (negbin.south east);
  \end{tikzpicture}
  \caption{Normal, Poisson, \& Uniform clusters}
  \label{fig:3}
\end{figure}

\begin{figure}
  \centering
  \begin{tikzpicture}
    \node[inner sep=0pt] (exp) {\includegraphics[width=\textwidth]{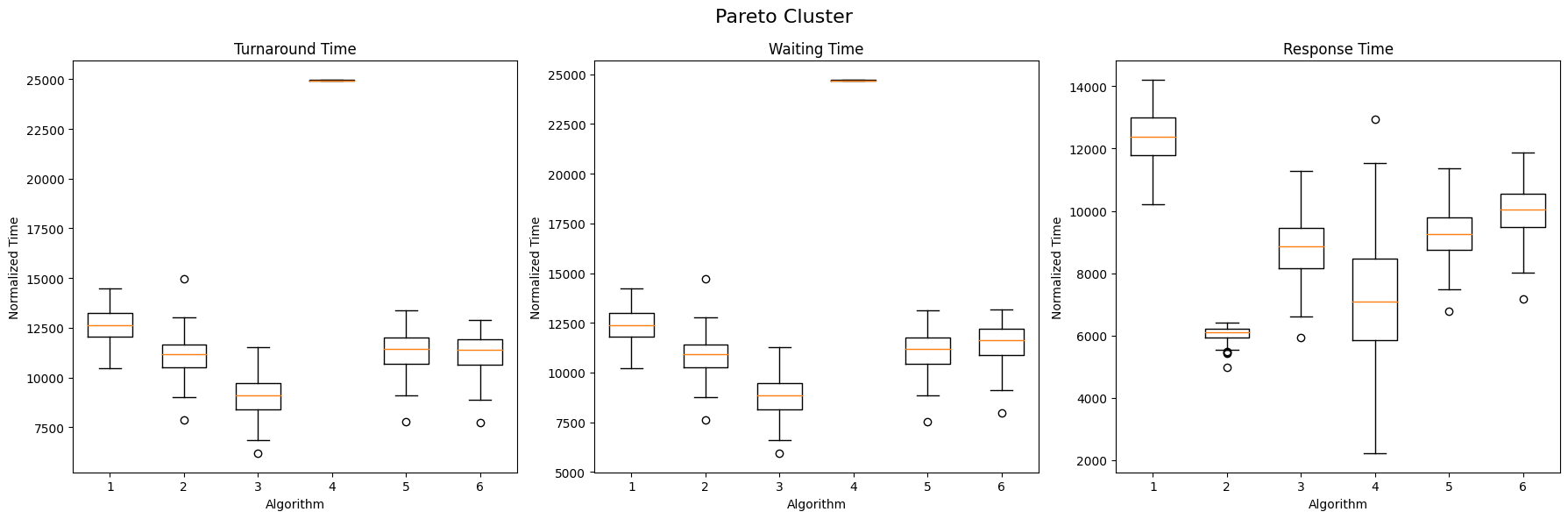}};
    \node[inner sep=0pt, below=0.2cm of exp] (geom) {\includegraphics[width=\textwidth]{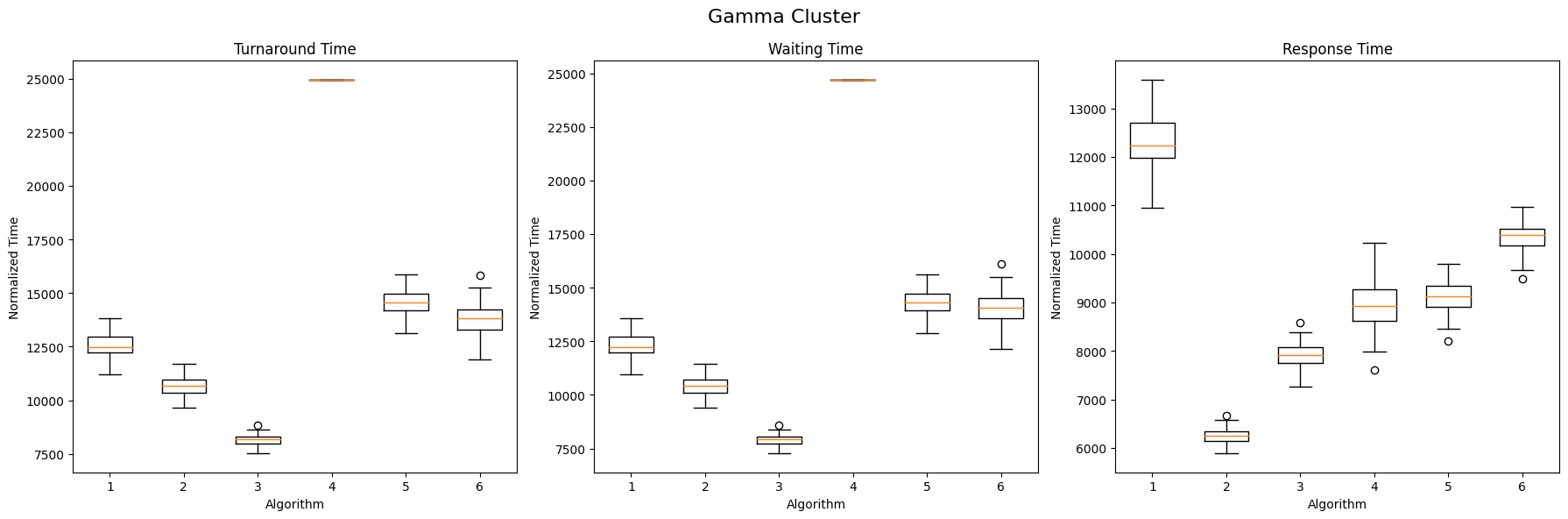}};
    \node[inner sep=0pt, below=0.2cm of geom] (negbin) {\includegraphics[width=\textwidth]{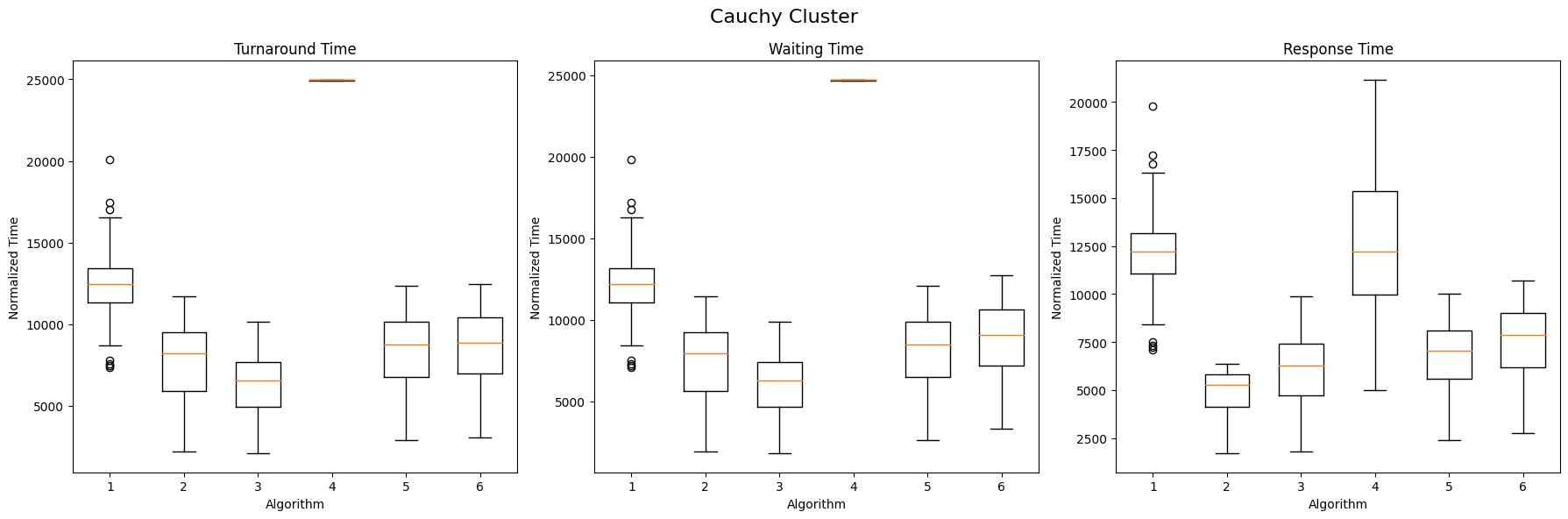}};
    \draw[black, line width=0.5pt] (exp.north west) rectangle (negbin.south east);
  \end{tikzpicture}
  \caption{Pareto, Gamma, \& Standard Cauchy (Lorentz) clusters}
  \label{fig:4}
\end{figure}

\begin{figure}
  \centering
  \begin{tikzpicture}
    \node[inner sep=0pt] (exp) {\includegraphics[width=\textwidth]{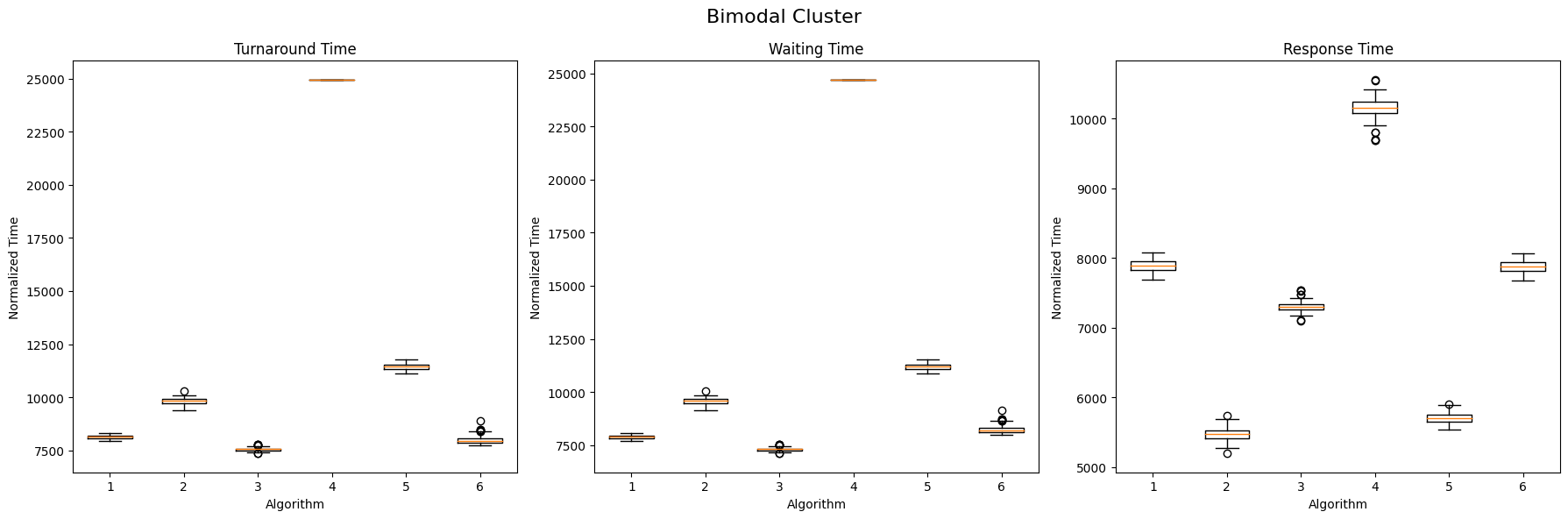}};
    \node[inner sep=0pt, below=0.2cm of exp] (geom) {\includegraphics[width=\textwidth]{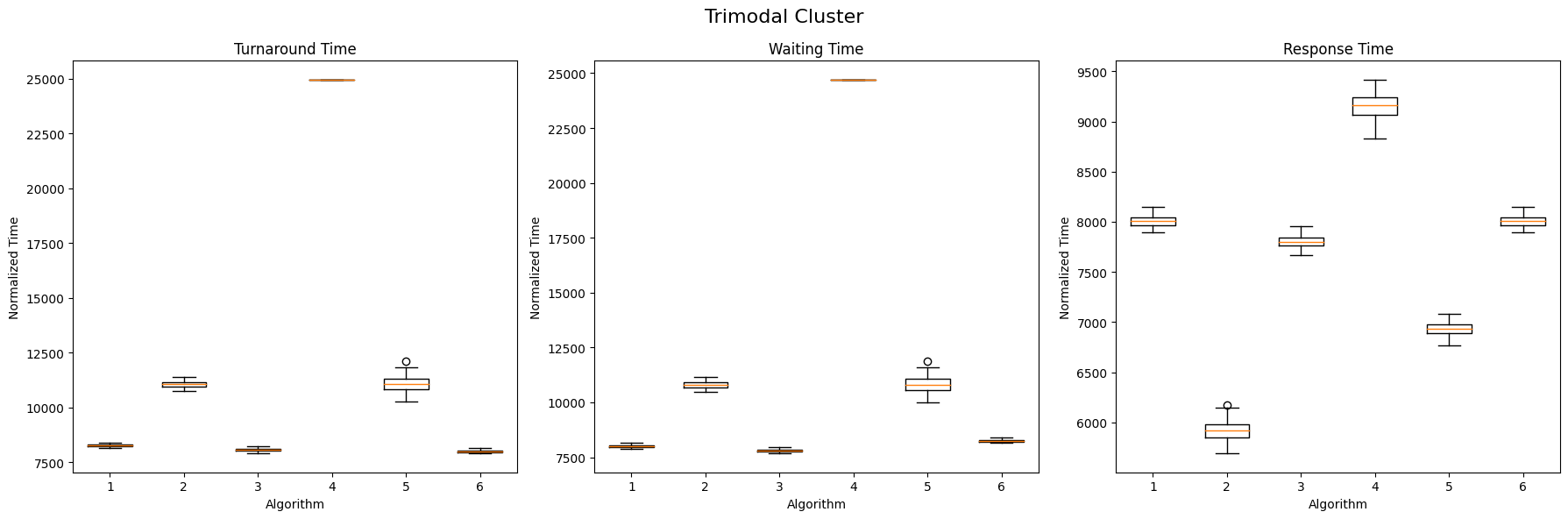}};
    \node[inner sep=0pt, below=0.2cm of geom] (negbin) {\includegraphics[width=\textwidth]{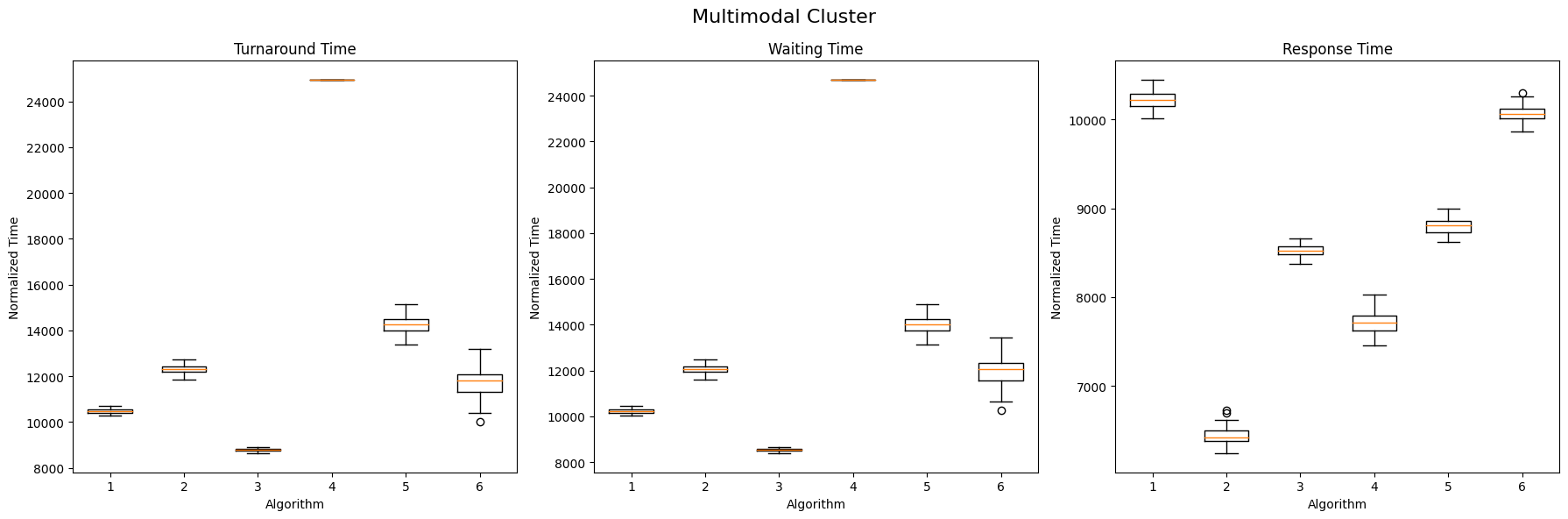}};
    \draw[black, line width=0.5pt] (exp.north west) rectangle (negbin.south east);
  \end{tikzpicture}
  \caption{Bimodal, Trimodal, \& Tultimodal clusters}
  \label{fig:5}
\end{figure}

\subsection{Description of the Test Cases}
 Unlike recent works  \cite{mora2017modified, shyam2014improved, hyytia2016round, xiuqin2012diffserv, banerjee2012comparative, mishra2014improved} and others where primarily evaluations of algorithms are based on mere set of unsystematic, tiny examples, we aim to provide a robust evaluation by considering diverse set of clusters of jobs and a large number of test cases. To the best of our knowledge, there is no publicly available, sufficiently large, domain-agnostic, benchmark dataset of \textit{bursttime}s that would be \textit{suitable} for comparing under our setting and methodologies. It was a significant motivation for us to first create a diverse and sufficiently large dataset of \textit{bursttime}s. We have made it publicly available with the aim of enabling the research community to utilize it in their own work. Eventually, we evaluated all algorithms on this dataset to assess their performance. For this evaluation, first of all we have carefully selected several  probability distributions to generate the test cases: we have considered  Normal, Exponential, Geometric, Negative Binomial, Poisson, Uniform, Pareto, Gamma, and Standard Cauchy (Lorentz) distribution. The Normal distribution enabled us to assess algorithms' ability to handle data clustering around a mean value, while the Exponential distribution shed light on performance in scenarios with exponential decay-like patterns. The Geometric distribution offered insights into algorithms' response to decreasing probabilities of longer burst times. Simulating long-tailed variations, the Negative Binomial distribution, being a general case of the geometric distribution, allowed us to scrutinize algorithm behavior in such scenarios. By employing the Poisson distribution, we observe algorithm behavior under specific occurrence patterns of processes. The Uniform distribution challenged algorithms with burst times exhibiting a uniform and evenly distributed pattern. The Pareto distribution, on the other hand, is characterized by a heavy tail, making it suitable for modeling phenomena where a small number of events have a disproportionately large impact. The Gamma distribution, with its shape and rate parameters, offers versatility in modeling various phenomena, including wait times and service times. The Standard Cauchy distribution, also known as the Lorentz distribution, represents a distribution with heavy tails and no defined mean or variance, providing insights into scenarios with extreme outliers. Together, these distributions provide a comprehensive overview of diverse \textit{bursttime} patterns \cite{ostle1963statistics}, offering a holistic assessment of algorithm performance across various real-world scenarios, which has not been covered in previous studies. While the distributions we've covered cover many scenarios, sometimes load on the CPUs can be even more varied and complex. To ensure our evaluations are robust, we've also considered bimodal, trimodal, and multimodal distributions. In a continues frame of reference (the CPU is processing arbitrarily large number of jobs),these distributions exhibit more irregular and diverse \textit{bursttime} patterns. In the first phase of analysis we study the first nine distributions and in the second phase we investigate into these three special distributions.
 
 For each distribution, we first produce 100 test cases, with each test case consisting of 100 integers \textit{bursttime}s.
 After this initial phase, we fine tune the entire dataset. While distributions are primarily generated\footnote{\url{https://numpy.org/doc/1.16/reference/routines.random.html}} by randomly sampling values, these samples may not align with the realistic characteristics of job burst times, where each job's burst time must fall within a feasible range. Simply relying on theoretical distributions could lead to unrealistic or unfeasible burst times, skewing the evaluation of scheduling algorithms. Our fine tuning are 2 phase:
 \begin{enumerate}
     \item  \textbf{Adhering to Practical Constraints}: While we initially generated instances from several distributions to emulate diverse loads on the CPU, we recognize that CPUs do not process loads exactly the way they arrive in the system (in batch form here) following a precise distribution. CPUs cannot process arbitrarily small portions of a job; they allocate a minimum execution time for all jobs by default. Similarly, CPUs cannot allocate arbitrarily large processes. If a process exceeds the CPU's allocation, the remainder of the process is stored in secondary memory. Using several page ranking algorithms \cite{4809246}, the CPU executes these remaining portions in subsequent iterations or batches \cite{silberschatz2006operating}. To account for this, we ensure that our dataset reflects this practical constraint, making it more applicable not only for rigorous evaluations but also for practical scenarios: After generating 100 samples for each distribution, we `clip' each sample to fall within the range of 1 to 500 for the \textit{bursttime}s of each job. This process reflects resource allocation constraints inherent to CPUs. Values less than 1 are set to 1, representing the minimum resource allocation that a CPU can assign to a job. Similarly, values greater than 500 are set to 500, representing the maximum resource allocation that a CPU can assign to a job. This is how, from a set of randomly sampled distributions, we finalized these job clusters, adhering to the realistic constraints of CPU processing.
     \item  \textbf{Normalisation of the Clusters}: After the clusters are ready, we normalise all finalised samples. We divide each \textit{bursttime} by the sum of the bursts present in the sample. After that we multiply each burst by 25000. We normalised each sample such that the sum of bursts in each sample across clusters are (almost) equal to 25000. By scaling the data, we eliminate the influence of magnitude disparities, allowing us to focus solely on the algorithmic performance with respect to \textit{bursttime} distribution patterns between any pair of samples. 

\end{enumerate}
Furthermore,  for a fine-grained comparison, we've used box plots for each metric to analysis across various distributions for each algorithm. We've referred to our algorithm as ``\textit{FairBatch}" here and have taken the ceil of mean of the \textit{bursttimes} as the \textit{timeQuantum} in RR as it remains almost equal for all the lists under examination.In diagram. We describe the legends used in experiments in table \ref{tab:clusters_algorithms}.

\begin{table}

\begin{tabular}{c|c|c}
        \hline
        Abbreviation & Description & Used in \\
        \hline
        Nor Cluster & Collection of arrays of jobs following the underlying Normal distribution & Fig:\ref{fig:2}, \ref{fig:3}, \ref{fig:4}, \ref{fig:5} \\
        Expo Cluster & Collection of arrays of jobs following the underlying Exponential distribution & Fig:\ref{fig:2}, \ref{fig:3}, \ref{fig:4}, \ref{fig:5} \\
        Geo Cluster & Collection of arrays of jobs following the underlying Geometric distribution & Fig:\ref{fig:2}, \ref{fig:3}, \ref{fig:4}, \ref{fig:5} \\
        N.Bio Cluster & Collection of arrays of jobs following the underlying Negative Binomial distribution & Fig:\ref{fig:2}, \ref{fig:3}, \ref{fig:4}, \ref{fig:5} \\
        Psn Cluster & Collection of arrays of jobs following the underlying Poisson distribution & Fig:\ref{fig:2}, \ref{fig:3}, \ref{fig:4}, \ref{fig:5} \\
        Ufm Cluster & Collection of arrays of jobs following the underlying Uniform distribution & Fig:\ref{fig:2}, \ref{fig:3}, \ref{fig:4}, \ref{fig:5} \\
        Bimodal Cluster & Collection of arrays of jobs following the underlying Bimodal distribution & Fig:\ref{fig:2}, \ref{fig:3}, \ref{fig:4}, \ref{fig:5} \\
        Trimodal Cluster & Collection of arrays of jobs following the underlying Trimodal distribution & Fig:\ref{fig:2}, \ref{fig:3}, \ref{fig:4}, \ref{fig:5} \\
        Multimodal Cluster & Collection of arrays of jobs following the underlying Multimodal distribution & Fig:\ref{fig:2}, \ref{fig:3}, \ref{fig:4}, \ref{fig:5} \\
        Gamma Cluster & Collection of arrays of jobs following the underlying Gamma distribution & Fig:\ref{fig:2}, \ref{fig:3}, \ref{fig:4}, \ref{fig:5} \\
        Cauchy Cluster & Collection of arrays of jobs following the underlying Cauchy (Lorentz) distribution & Fig:\ref{fig:2}, \ref{fig:3}, \ref{fig:4}, \ref{fig:5} \\
        Pareto Cluster & Collection of arrays of jobs following the underlying Pareto distribution & Fig:\ref{fig:2}, \ref{fig:3}, \ref{fig:4}, \ref{fig:5} \\
        \hline
        1 & FCFS & Fig:\ref{fig:2}, \ref{fig:3}, \ref{fig:4}, \ref{fig:5} \\
        2 & FairBatch & Fig:\ref{fig:2}, \ref{fig:3}, \ref{fig:4}, \ref{fig:5} \\
        3 & SRTF & Fig:\ref{fig:2}, \ref{fig:3}, \ref{fig:4}, \ref{fig:5} \\
        4 & LRTF & Fig:\ref{fig:2}, \ref{fig:3}, \ref{fig:4}, \ref{fig:5} \\
        5 & RR & Fig:\ref{fig:2}, \ref{fig:3}, \ref{fig:4}, \ref{fig:5} \\
        6 & CFS (SCHED\_BATCH) & Fig:\ref{fig:2}, \ref{fig:3}, \ref{fig:4}, \ref{fig:5} \\
        \hline
    \end{tabular}
    \caption{Abbreviations and descriptions of clusters and algorithms}
    \label{tab:clusters_algorithms}
\end{table}

\subsection{Discussion}
Each algorithm is tested on 100 test cases present in each cluster, we represent their aggregated results in the form of boxplots. Before analysis, we are presenting a concise introduction of boxplots for readers at appendix B \ref{B}.\newline
\begin{itemize}
    \item \textbf{First Phase of the Analysis}: In this phase, we investigate the algorithms against
    `Nor cluster', `Expo cluster', `Geo cluster', `N.Bio cluster', `Psn cluster', `Ufm cluster', `Gamma cluster', `Cauchy cluster', `Pareto cluster'. These comparisons are primarily made to simulate homogeneous yet diverse patterns of CPU loads. For our average turnaround and waiting time, the benchmarking algorithm is SRTF as mentioned. We compare the algorithms based on their comparative performance with respect to SRTF as demonstrated across \ref{fig:2},\ref{fig:3},\ref{fig:4}. We ignore the outliers and relatively compare the performances based on the IQR, median, and range. 

    If we discretely compare the average turnaround, waiting and response time for FairBatch (2 in the figures):
    \begin{enumerate}
        \item Out of \textbf{9 cases}, FairBatch has achieved the least average \textit{turnaroundtime} in \textbf{7 cases} after SRTF. In the case of `Nor Cluster', FairBatch has achieved the third least average \textit{turnaroundtime} after FCFS (1 in figures) and SRTF. In the case of `Psn Cluster', it comes after SRTF, FCFS, and SRTF respectively in \ref{fig:3}, 
    
        \item Out of \textbf{9 cases}, FairBatch has achieved the least average \textit{waitingtime} in \textbf{7 cases} after SRTF. In the case of `Nor Cluster' in \ref{fig:3}, FairBatch has achieved the third least average \textit{waitingtime} after FCFS (1 in figures) and SRTF.  In the case of `Psn Cluster', it comes after SRTF, FCFS, and SRTF respectively in \ref{fig:3}.

        \item Out of \textbf{9 cases}, FairBatch has achieved the least average \textit{responsetime} in \textbf{7 cases}. In the case of `Nor Cluster' and `Psn Cluster' in \ref{fig:3}, FairBatch has achieved the second least average \textit{responsetime} after LRTF (4 in figures). The only algorithm that ever exceeds FairBatch in \textit{responsetime} is LRTF only in these 2 cases.
    \end{enumerate}

\textsc{Investigating the performance in the case of `Nor Cluster'}\newline

Here, the underlying distribution is the normal distribution, where approximately 68\% of the data falls within one standard deviation of the mean, about 95\% falls within two standard deviations, and around 99.7\% falls within three standard deviations of the mean  \cite{ostle1963statistics}. So, in this cluster, the majority of the jobs' burst times are centred around a single value (the mean). In other words, most jobs are very comparable in size (\textit{bursttime}s). Now, FCFS, being a non-preemptive algorithm, completes jobs without any context switch. As a result, here in terms of turnaround and waiting time, FCFS excels. However, due to FCFS's non-preemptive nature, it cannot respond to other jobs waiting in the queue before one job is completely executed, leading to the issue of the convoy effect  \cite{silberschatz2006operating}. FairBatch, in this case, preemptively switches contexts based on the fairness ratio, which prioritizes \textit{waitingtime} for jobs as well as \textit{progress} in currently under-executing jobs. This is evident when comparing the average \textit{responsetime} of FairBatch and FCFS. FairBatch not only competitively prioritizes efficiency but also gives weightage to fairness in job selection.

On the other hand, LRTF achieves the best \textit{responsetime} in this cluster. LRTF, being a preemptive algorithm, at every instance chooses the longest available job, performs bare minimum execution, and moves on to the next largest job. As most of the jobs are similar in size, this procedure results in excessive preemption, providing a ``response" to each job in the least time. While FCFS was completely focused on efficiently executing all the jobs one after another, causing severe convey in the system, LRTF aggressively responded to almost every available job with marginal progress, which drastically reduced its efficiency \ref{fig:3}. FairBatch, on the other hand, strikes an excellent balance between efficiency and fairness that is unparalleled even in SRTF.\newline

\textsc{Observing the trade-off Arose in the case of `Psn cluster'}\newline

The Poisson distribution is primarily used to model the likelihood of an event occurring a certain number of times within a specific period. We have selected this discrete distribution to examine the occurrence of repetitive loads in CPUs  \cite{mishra2020simulation}. Since batch processing is a CPU-bound process, repetitive loads are frequently observed in the system. Consequently, this cluster is heavily influenced by frequently occurring comparable jobs. FCFS, as a non-preemptive algorithm, aggressively executes repetitive jobs, causing other jobs to wait disproportionately and eventually starve. The next algorithm that outperforms FairBatch is CFS. As CFS (SCHED\_BATCH) is a widely used finely tuned scheduling policy for large batches, operating on the principles of virtual runtime and nice values, it is well-suited for managing commonly observed repetitive loads on systems. However, due to its design\footnote{\url{https://elixir.bootlin.com/linux/v5.19.9/source/kernel/sched/fair.c##L7429}} lacking a suitable preemption-supportive mechanism and heuristics, it does not perform well in terms of response time.

On the other hand, LRTF aggressively preempts and responds to frequently repetitive loads. Similar to the `Nor Cluster', while it achieves excellent responsiveness, it shows marginal efficiency. In contrast, FairBatch does not monopolize resources for repetitive jobs but evenly distributes them among all jobs while maintaining competitive efficiency. This demonstrates that the FairBatch algorithm exhibits strong adaptability and resilience in handling the randomness present in different data distributions.
\item \textbf{Second Phase of Analysis}: In the previous phase, we have observed the unparalleled superiority in terms of both efficiency and fairness of FairBatch across 7 out of 9 cases. We have also analysed when there is a harsh trade-off between efficiency and fairness, how all algorithm falls apart in consistently maintaining the trade-off except FairBatch. In order to delve into more on these scenarios comprising of critical trade-offs, after the first phase, here we primarily investigate bimodal, trimodal and multimodal clusters.
    If we discretely compare the average turnaround, waiting and response time for FairBatch in \ref{fig:5}:

\begin{enumerate}
    \item FairBatch achieves the least average \textit{turnaroundtime} after FCFS and SRTF in all three cases.
    \item FairBatch achieves the least average \textit{waitingtime} after FCFS and SRTF in all three cases.
    \item FairBatch achieves the least average \textit{responsetime} in all three cases.
\end{enumerate}
In regular batch processing, heterogeneous loads are not commonly observed unless the batch is sufficiently large. These three distributions were created by concatenating more than one heterogeneous subpopulation of normal distributions. As a result, the central tendency of the measurements is inherently more complex than the previous clusters. These clusters are generated to adversarially test the algorithms to observe how they perform when there are multiple distinct subpopulations creating heterogeneous loads on the CPU. Depending on the sequence of the subpopulations, the arrival sequence of the jobs changes, potentially resulting in a change in the average turnaround, waiting, and response times.

FCFS has a distinct advantage in these cases: it minimizes the maximum \textit{turnaroundtime} across jobs for any finite arrival sequence of jobs  \cite{grosof2021nudge}. Irrespective of the sequence of these subpopulations, FCFS achieves the least turnaround and waiting times across all three clusters after SRTF in our experiment. While it achieves superior efficiency due to its theoretical guarantees, it is evident from Figure \ref{fig:5} that it does not perform well in terms of \textit{responsetime} in any of these clusters. It suffers from severe starvation and lags in fairness in job selection. FairBatch in all these cases performs excellently in handling the efficiency and fairness trade-off as evident from the results in figure \ref{fig:5}. Not only FCFS, but also CFS and LRTF along with others do not perform well when it comes to selecting jobs in a fairer way while maintaining competitive efficiency unlike FairBatch.\newline

\item \textsc{Stability of the performance: another advantage of FairBatch}\newline

FairBatch offers another distinct advantage over algorithms like Round Robin and FCFS due to its stability and predictability. While the sequence of execution and overall results of these algorithms heavily depend on the sequence in which the job arrived, FairBatch eliminates this variability by arranging and executing jobs based on \textit{fairnessratio} and \textit{timeQuantum}. This characteristic makes the algorithm a more reliable and consistent scheduling algorithm, ensuring stability and predictability in its performance.\newline
\end{itemize}

After a comprehensive comparison of all algorithms across various distributions as demonstrated in fig:\ref{fig:2}, \ref{fig:3}, \ref{fig:4}, \ref{fig:5}, it is evident that the FairBatch performs exceptionally competitively with respect to the benchmark (SRTF) for all distributions showcasing its superiority through the delicate balance it strikes between drastically reducing \textit{waitingtime}, ensuring responsiveness, and eliminating starvation. With stability, predictability, and unmatched efficiency, our algorithm bridges the gap between suitable responsiveness and required efficiency through a fair and holistic approach, as demonstrated extensively.

\section{Conclusion}
Our research paper addresses a significant gap in the scheduling of batch processing systems by leveraging the favourable aspects of advanced algorithms. While classical algorithms like SRTF, RR, and LRTF, CFS etc have been predominantly utilised in interactive systems, their potential benefits have been mostly overlooked in the context of batch processing. Our proposed algorithm aims to bridge this gap by harnessing the positive attributes of these advanced algorithms without inheriting their limitations in a unique and wholesome way. By focusing on fairness, efficiency, and system performance, our algorithm provides a comprehensive solution that outperforms traditional approaches in multiple parameters. Through careful consideration of the strengths and weaknesses of existing algorithms, we have developed a novel approach that optimises scheduling in batch systems, effectively addressing the existing void and achieving superior results.

\section*{Data Availability}
The data that support the findings of this study are openly available at \href{https://github.com/humblef-oo-l/scheduling-test-cases/blob/0323937576c0ccc4391722bf7252fe6da60e5613/testCases.pdf}{GitHub}
\section*{Conflict of Interest}
On behalf of all authors, the corresponding author states that there is no conflict of interest.
\section*{Acknowledgement}
The authors express their gratitude to the reviewers, editor, and journal administrator for their cooperation. Special thanks to John Schroder for his unwavering assistance, thoughtful insights, and moral support throughout the project.

\bibliography{snb}

\newpage
\section*{Appendix A}\label{A}

Commonly used scheduling polices are hereby discussed:
\begin{itemize}
\item First-Come, First-Served (FCFS):
The FCFS algorithm, a non-preemptive scheduling algorithm, operates on a simple principle: the process that arrives first is served first. This algorithm is easy to implement and ensures fairness in process execution. However, it suffers from poor average response time and can lead to a phenomenon known as the ``convoy effect" where long processes hold up shorter ones.

\item Shortest Job Next (SJN/SPT) - Non-Preemptive and Preemptive Versions:
The SJN algorithm, also referred to as the shortest job first algorithm, aims to minimise the average waiting time of processes. In its non-preemptive version, it selects the process with the smallest total execution time (burst time) for execution. While, it is well known that for single server systems without preemption, SJN provides optimal performance in terms of minimising \textit{waitingtime}  \cite{conway1967theory, silberschatz2006operating}.

In the preemptive version of SJN, known as Shortest Remaining Time Next (SRTF), the algorithm allows for the preemption of the currently executing process if a shorter job enters the system. This ensures that shorter jobs are given higher priority, leading to improved response time. This algorithm serves as a benchmark for average waiting and turnaround time However, SRTF may introduce starvation and additional overhead due to frequent context switching.

\item Round Robin (RR):
The Round Robin algorithm addresses the limitations of FCFS by introducing time slicing. It allocates a fixed \textit{timeQunatum} to each process in a cyclic manner. If a process does not complete within its \textit{timeQunatum}, it is preempted and moved to the end of the queue. RR ensures fairness among processes and prevents starvation. However, it may lead to high overhead due to frequent context switching.

\item Priority Scheduling - Non-Preemptive and Preemptive Versions:
Priority Scheduling assigns priority levels to processes based on their characteristics, such as the importance of the task, deadline constraints, or resource requirements. In the non-preemptive version, the highest priority process is scheduled and allowed to run until completion. This approach can lead to higher-priority processes monopolising the CPU and potentially starving lower-priority processes.

The preemptive Priority Scheduling allows for the preemption of the currently executing process if a higher-priority process becomes available. This ensures that processes with higher priority are promptly executed, maintaining fairness and responsiveness. However, frequent preemption introduces additional overhead and can impact system performance.

\item Longest Remaining Time First (LRTF/LRPT): is a preemptive CPU scheduling algorithm that selects the process with the longest remaining execution time. It allocates the CPU to the process with the greatest time requirement from the available processes. Preemption is permitted, allowing higher-priority processes with longer execution times to interrupt the currently running process. The non-preemptive version is called LPT or Longest Processing Time Rule which is the opposite policy of SJF and maximises the quantities SJF minimises  \cite{conway1967theory}. 

\item Multilevel Queue and Multilevel Feedback Queue Scheduling,  \cite{harki2020cpu}:
Multilevel Queue Scheduling categorises processes into multiple priority queues, each with its own scheduling algorithm. Processes are initially assigned to a specific queue based on their attributes, such as priority or type. This approach provides a balance between fairness and responsiveness, as different classes of processes receive appropriate treatment. However, determining the optimal number and configuration of queues can be challenging. Microsoft's Windows Operating system, by default,  employs Multilevel feedback queue scheduling along with a subset of FCFS, RR, SJF/SRTF and Priority Scheduling  \cite{krzyzanowski2015process}.
\item Completely Fair Scheduler(CFS): This is the classical scheduling policy of several Linux distributions  \cite{kernelSchedulerx2014}. It implements a classic scheduling method called weighted fair queuing  \cite{li2009efficient,nagle1987packet}, which originated from packet networks and was earlier applied to CPU scheduling under the name stride scheduling  \cite{waldspurger1995lottery}. Unlike traditional policies, which prioritize tasks based on their arrival or execution time, CFS uses a concept called ``virtual runtime" to determine the order of task execution. Each task is assigned a ``nice value," which influences its priority. Tasks with lower nice values (higher priority) are given more virtual runtime compared to tasks with higher nice values (lower priority). This ensures that tasks are scheduled fairly, providing a more balanced allocation of CPU time among competing processes  \cite{kernelSchedulerx2014}. Unlike previous implementations, CFS is notable for being the first widespread adoption of fair queuing in a general-purpose operating system as a process scheduler.
\end{itemize}
\newpage
\section*{Appendix B}\label{B}
\begin{figure}[h]
  \centering
  \begin{tikzpicture}
    \node[inner sep=0pt] (exp) {\includegraphics[width=0.3\textwidth]{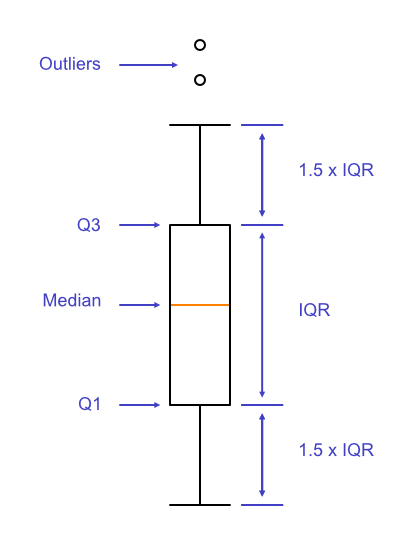}};
    \draw[black, line width=0.3pt] (exp.north west) rectangle (exp.south east);
    \end{tikzpicture}
    \caption{Boxplot}
    \label{Boxplot Description}
\end{figure}

\textbf{Box plots} (Fig:\ref{Boxplot Description}) are data visualization techniques that provide a concise summary of the distribution of a dataset. They consist of a box, which represents the interquartile range (IQR) containing the middle 50\% of the data, with a line inside marking the median. Whiskers extend from the box to the minimum and maximum values within a certain range, often 1.5 times the IQR. Outliers beyond the whiskers are plotted individually.

\begin{itemize}[leftmargin=*]
    \item \textbf{Median (Q2):} Middle value of the dataset.
    \item \textbf{Upper Quartile (Q3):} Median of the upper half of the dataset.
    \item \textbf{Lower Quartile (Q1):} Median of the lower half of the dataset.
    \item \textbf{Interquartile Range (IQR):} Range containing the middle 50\% of the data.
    \item \textbf{Upper Whisker:} Maximum value within 1.5 times the IQR from Q3.
    \item \textbf{Lower Whisker:} Minimum value within 1.5 times the IQR from Q1.
    \item \textbf{Outliers:} Individual data points beyond the whiskers.
\end{itemize}
\textbf{Interpreting Boxplots}:\newline

In performance analysis, the common practice is to focus on the median, IQR, and range  \cite{abedin2014r,frigge1989some}:

\begin{enumerate}

  \item \textbf{Median}: Less sensitive to outliers compared to the mean, providing a more reliable measure of central tendency.
  
  \item \textbf{IQR}: A robust measure of the spread of the middle 50\% of the data, less influenced by extreme values.
  
  \item \textbf{Range}: Gives an indication of the overall spread of values in the dataset, despite being affected by outliers.

    \item\textbf{Ignoring Outliers}: While outliers can offer insights into extreme data points, they can also distort the analysis. By focusing on the IQR, or median, we can mitigate the impact of outliers and obtain a more accurate understanding of the typical behavior of the data  \cite{abedin2014r}.

\end{enumerate}
\end{document}